\documentclass{elsart}
\usepackage{graphics}

\def\beq\begin{equation}
\def\eeq\begin{equation}
\def\bea\begin{eqnarray}
\def\bea\end{eqnarray}
\begin{document}
\begin{frontmatter}



\title{
Buckyballs and gluon junction networks on the femtometer scale\thanksref{*}
}
\thanks[*]{Dedicated to J. Zim\'anyi on the occasion of his 70th birthday.}

\author[label1,label2,label3]{T. Cs\"{o}rg\H{o}}
\ead{csorgo@sunserv.kfki.hu}
\author[label3,label4]{M. Gyulassy}
\ead{gyulassy@mail-cunuke.phys.columbia.edu}
\author[label5]{D. Kharzeev}
\ead{kharzeev@bnl.gov}

\address[label1]{MTA KFKI RMKI, H-1525 Budapest 114, POB 49, Hungary }
\address[label2]{Inst. de Fisica, USP,
	CP 66318, 05389-970 Sao Paulo, Brazil}
\address[label3]{Dept. Physics,
	Columbia University, 538W 120th St, New York, NY 10027, USA }
\address[label4]{
	Collegium Budapest, Szenth\'aroms\'ag u. 2, H-1014 Budapest, Hungary }
\address[label5]{Dept. Physics, Brookhaven National Laboratory,
	Upton, NY 11973-5000, USA }

\begin{abstract}
We explore the possibility that novel geometrical structures analogous to
carbon Fullerenes may exist in Nature on the femtometer scale.  
The theory of strong interactions, Quantum Chromo Dynamics (QCD)
predicts the existence of special topological gluon field configurations 
called baryon junctions and anti-junctions. 
Here we show that femto-scale structures, networks or closed
(gluon field) cages, can be constructed in the theory of QCD as  tiny 
cousins of familiar nano-scale structures such as carbonic Fullerenes 
$C_{60}$, $C_{70}$.
The most symmetric polyhedra of QCD junctions  (J-balls)  are characterized by
the ``magic numbers'' 8, 24, 48, and 120, and zero net baryon number. Tubes,
prisms, tori and other topological structures can also be created. In addition,
special configurations can be constructed that are odd under charge and parity
conjugation (CP),  although the QCD Lagrangian is CP even. 
We provide a semi-classical estimate for the
expected mass range of QCD Buckyballs and discuss the possible conditions
under which such novel topological  excitations of the QCD vacuum may be
produced   in  experiments of high energy physics.
\end{abstract}

\end{frontmatter}

\section{Introduction}
\label{s:int}
\begin{figure}
\vspace*{10.5cm}
\includegraphics{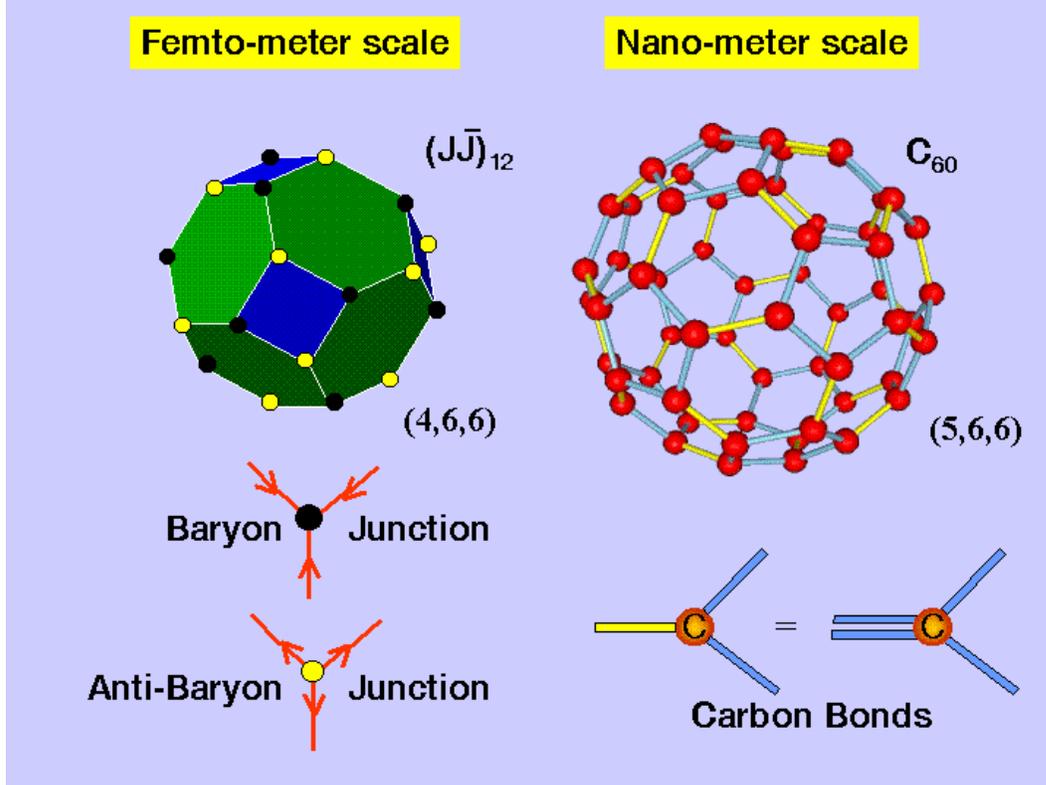}
\caption{Femtometer scale QCD analogs of  nanometer scale( QED) carbon Fullerenes
and their  corresponding three way junction building blocks.}
\label{f:fig1}
\end{figure}

The Buckyball is the nickname for the carbon molecule Buckminsterfullerene,
$C_{60}$, a new form of carbon discovered in chemistry in 1985 by R. F. Curl, H. W.
Kroto and R. E. Smalley\cite{jb1}. The molecule was named after the geodesic dome,
invented by the architect Buckminster Fuller, whose geometry approximates that
of a truncated icosahedral (soccer-ball) shaped structure. The discovery of
Buckyballs was followed by the discovery of a wide variety of other carbon
molecules with interesting geometrical properties. Carbon tubes, helixes, tori,
etc. opened the doorway to technology on the nanometer ($10^{-9}$ m) scale. Carbon
atoms can be arranged in novel geometric forms because the carbonic bonds can
arrange into 3 way junction structures as illustrated in Fig.~\ref{f:fig1}.
Nano-structures have also been constructed  using 3 and 4 way DNA junctions by
Seaman et al.~\cite{jb2} . The field of nano-technology is developing rapidly  using an
assortment of  molecular junctions as the chemical ``lego" building blocks.

In nuclear/particle physics, where the
distance scales are  femtometers  ($10^{-15}$ m),  the existence of  special
three-way  QCD junctions (topological gluon field configurations) was predicted
 a long time ago\cite{jb3} . Lattice QCD calculations were able to confirm the existence
of such junctions only recently\cite{jb4}.  Data on baryon stopping and strangeness
production in experiments with high energy heavy ion collisions from CERN SPS
and  BNL RHIC accelerators are also in agreement with model calculation
assuming that  QCD junctions carry the conserved baryon charge\cite{jb5,jb6,jb7,jb8}. 
In this Letter, we explore what types of  femtometer scale structures can  be
constructed from QCD using junctions and anti-junctions as a nuclear scale
``lego set".  Our preliminary results were presented at a Symposium on
multiparticle production in high energy physics~\cite{jb9}.

According to QCD, hadrons are composite bound state configurations built up
from the fundamental quark and gluon fields. Quarks, $\Psi_{i,f}(x)$, 
carry color, $i=1, ... , N_c$ , and
flavor,  $f=u,d,s,c,b,t$  quantum numbers. Gluons, $A_a^\mu(x)$, 
are the vector gauge bosons intermediating the color,
$a = 1, ... , N_c^2-1$, interactions between the quarks and gluons. The
form of the interaction is fixed by the principle of gauge invariance under the
non-Abelian color $SU(N_c)$  Lie group.  The $N_c=1$ limit is Quantum Electrodynamics (QED).
Gauge invariance of composite operators can only be achieved with the help of
open string operators, called Wilson lines~\cite{jb10}, that keep track of the phase
along an arbitrary path, $\Gamma$, in space-time. In QED, $U(\Gamma) =
\exp\left[i e \int_\Gamma dx^\mu A_\mu(x) \right]$  is the well known
Aharonov-Bohm phase~\cite{jb11} accumulated by an electron moving along a path $\Gamma$  
in an external electromagnetic field, $A^\mu(x)$ .   In QCD, $N_c = 3$ and
$U(\Gamma)$ is a matrix defined by a path
ordered exponential with dimension corresponding to that of the representation
of the generators, $T_a$, of the Lie algebra.

Closed Wilson loops, $\mbox{\bf Tr} U(\Gamma_{xx})$, 
correspond to color singlet 
glueball configurations in QCD, while open ``strings",
$\overline{\Psi}_{i_1f_1}(x) U^{i_1i_2}(\Gamma_{xy}) \Psi_{i_2f_2}(y)$,
 terminating with quark and anti-quark ends,
correspond to mesons.  Baryons are special field configurations composed of $N_c$
quarks with their  color flux strings tied together  (outer product of color
indices) by the  Levi-Civita antisymmetric tensor,
$\epsilon_{i_1 ... i_{N_c}}$ . In the physical  ($N_c = 3$) case,
baryons of flavor $(f_1,f_2,f_3)$ are represented 
by the color neutral and gauge invariant
operator,
\begin{equation}
B_{f_1f_2f_3} = 
	\overline{\Psi}_{i_1f_1}(x_1) \overline{\Psi}_{i_2f_2}(x_2) 
	\overline{\Psi}_{i_3f_3}(x_3) J^{i_1i_2i_3}(\Gamma_1,\Gamma_2,\Gamma_3),
\end{equation}
where the quark color indices are contracted by the baryon Junction tensor
\begin{equation}
J^{i_1i_2i_3}(\Gamma_1,\Gamma_2,\Gamma_3) = \epsilon_{j_1j_2j_3} 
	U^{i_1j_1}(\Gamma_1) U^{i_2j_2}(\Gamma_2)
	U^{i_3j_3}(\Gamma_3),
\end{equation}
that depends on the paths, $\Gamma_i$, 
connecting the quark at $x_i$ to an intermediate
junction vertex point, $x$. 
All three paths  are chromo-field flux lines oriented
into the junction vertex as represented by black dots in Fig. ~\ref{f:fig1}.
 Anti-baryons can be constructed similarly with 
the help of an anti-Junction tensor, $\overline{J}$, where
all the flux lines are oriented away from the vertex. Note that because  of the
special, $\det U = 1$, constraint on the symmetry group , $SU(3)$, 
$J^{i_1i_2i_3}(\Gamma,\Gamma,\Gamma)=1$ . Thus color singlet states can
be constructed from the color tensor links $U(\Gamma)$ 
not only by tracing and contracting with quark fields but also
 by contracting with baryon junctions.  The paths
from a physical junction vertex must be all nondegenerate. Paths are deformable
according to Stoke's theorem only if the background fields are pure gauge
artifacts. In the physical, confining vacuum, or in a quark-gluon plasma,
different paths correspond to configurations with different energy. In the
ground state of a heavy quark baryon, the physical junction vertex ends up in
the three quark plane, leading to a Y shaped chromo-field flux field
configuration inside the baryon~\cite{jb4}.

\section{Archimedean polyhedra in QCD}
\label{s:arch}
The compelling theoretical arguments in favor of the existence of gauge 
junction and anti-junctions as inevitable components of the Standard Model led
to the prediction~\cite{jb12} of 
$M^0_J = \mbox{\bf Tr} J\overline{J} = \epsilon U(\Gamma_1) U(\Gamma_2) U(\Gamma_3) \epsilon$, 
a new family of glueballs, 
 with masses {\it O}$(N_c)$ larger than usual glueballs corresponding
 to a closed string.  In addition, many new exotic
states formed by a multitude of quarks and anti-quarks~\cite{jb3,jb6} 
were predicted to exist.  So far none of these structures have been observed experimentally,
probably because the decay  widths of these structures is too  large, due to
their strong coupling to light meson and baryon anti-baryon states.  These
previously discussed QCD structures are analogous to carbonic structures with
low number of carbon atoms that do not possess any special geometric symmetry.

	In high-energy baryon and nuclear collisions, the valence quarks carry a large
fraction of the incident baryon's momentum. Those quarks  thus  hadronize in
the fragmentation regions which are typically  within one unit of rapidity, 
$y = 0.5 \log[(E+p_z)/(E-p_z)]$,
from the kinematic limits.  However, baryon junctions invalidate this naive
picture of baryon production, since gluons carry on the average only small
fraction of the baryon's momentum. Therefore, junction mechanism of  baryon
production (via exchange of the $M_J^0$ Regge trajectory) predicts a much higher
probability of finding the conserved valence baryon number many units of
rapidity away from the incident baryons~\cite{jb6}. In addition, 
junction dynamics also  naturally predicts~\cite{jb7} 
a high probability that the valence baryon  emerges with multiple
strangeness quantum numbers, e.g. $\Xi^-(dss)$, $\Omega^-(sss)$, 
in the central rapidity region, 
since the final baryon is made by neutralizing the color of  
the gluon junction by pair production of
quarks and antiquarks with arbitrary flavors.  The baryon
production data from SPS/CERN~\cite{jb7} and now RHIC/BNL~\cite{jb8} 
are consistent with these predictions 
and therefore lend experimental support to the important role that
gluon junction dynamics  plays in nuclear reactions.
%
%
From a rehadronizing quark matter baryon junctions may pick up the
valence quarks similarly as described by the quark combinatorics of the
ALCOR model that describes the production of multi-strange
anti-baryon to baryon ratios at CERN SPS in simple terms~\cite{alcor00}.
The success of the  ALCOR model implementation of  quark combinatorics 
in predicting~\cite{zim99} the multistrange
anti-baryon to baryon ratios at RHIC is thus consistent with  
a junction mechanism for the formation of baryons.

        Motivated  by Fullerenes,  in this Letter we point out the existence of
new geometric structures in QCD with  high spatial symmetry.  We determine the
geometric structure and the characteristic ``magic numbers" of these
configurations, using analogies with carbon Fullerene structures. We  explore
some of the interesting  topological structures that can be created by QCD
networks and  closed cages that may be produced in high energy nuclear
reactions joining multiple QCD junctions and anti-junctions. Although the QCD
Lagrangian  is CP even,  we point out that the junction and anti-junction
building blocks can be used construct  CP odd configurations that may also
serve as domain walls between inequivalent ($\theta$) QCD vacua.

 In QCD,  the orientation of flux lines going into (out of  anti-) junctions
restricts the set of allowed configurations. In particular, the number of
junctions has to be equal with the number of anti-junctions on any closed path
formed by the Wilson lines, which implies that QCD Fullerenes may have only
even number of vertexes V, and a zero net baryon number.
Recalling Euler's formula,
the number of faces ($F$),  the number of edges ($E$) and the number of vertices
($V$) of a simple (genus 0) polyhedron is related by  
\begin{equation}
V + F = 2 + E. 
\end{equation}
Since each
edge is sandwiched between a junction and anti-junction, each face must have an
even number of edges.  The number of faces,  
\begin{equation}
F = N_4 + N_6 + N_8 + ...,
\end{equation}
 is then a
sum of the the number of squares ($N_4$),  hexagons ($N_6$), etc. Each edge belongs
to two faces :  
\begin{equation}
 E  = (4 N_4 + 6 N_6 + 8 N_8 + ... )/ 2,
\end{equation} 
and  each vertex belongs to
three faces:  
\begin{equation}
 V  = (4 N_4 + 6 N_6 + 8 N_8 + ... )/ 3.   
\end{equation}
The resulting Diophantic
equations are solved by {\it any number of hexagons} and 
\begin{equation}
N_4 - \sum_{i=4}^\infty (i-3) N_{2i} = 6. 
\end{equation}
This implies that there is an infinite variety of Fullerene type 
of structures in QCD, similarly to the
case of carbon Fullerenes.

\begin{figure}
\vspace*{10.5cm}
\includegraphics{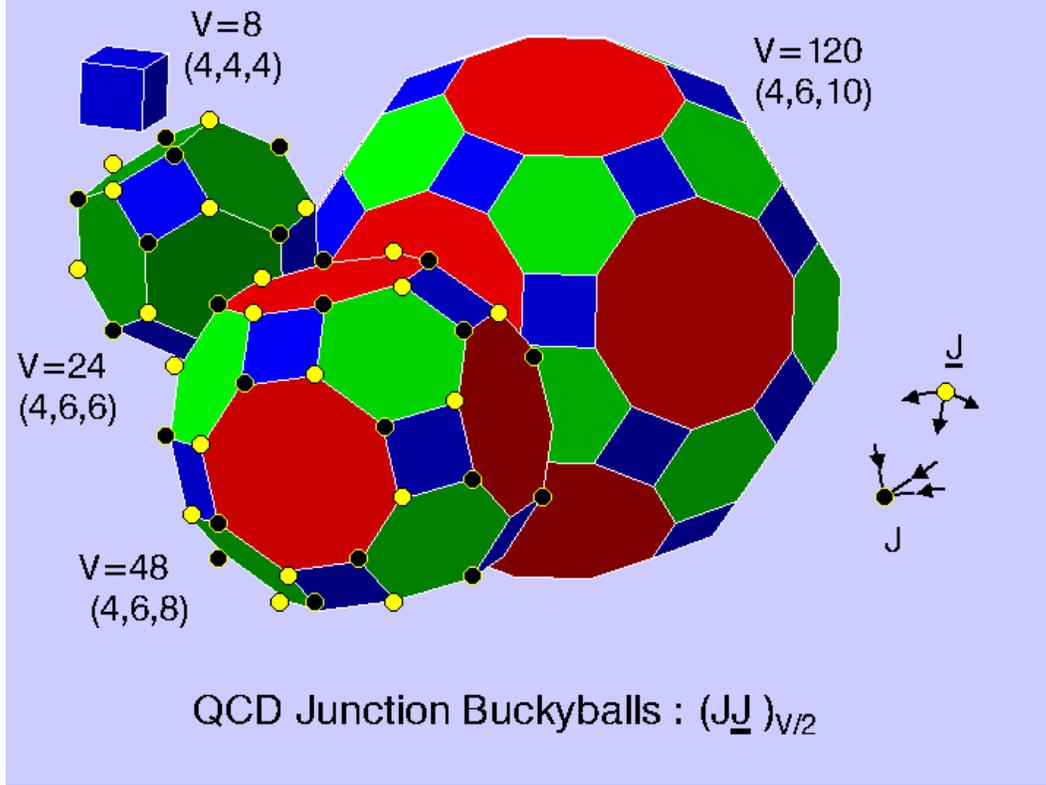}
\caption{The family of QCD Fullerenes, $(J\underline{J})_{V/2}$, 
with magic numbers $V=8$, $24$, $48$ and $120$.}
\label{f:fig2}
\end{figure}

We are particularly interested in the most symmetric geometric structures in
QCD, based on the expectation that configurations with the highest geometric
symmetry are the most stable ones, similarly to the case of the carbon
Fullerenes. If we require that all the vertex positions are equivalent with
each other, we have to find the so called Archimedean polyhedra with the
constraint that all faces have even number of edges. Archimedean polyhedra can
be characterized by the number of vertexes or, equivalently, by their vertex
structure $(i,j,k)$ denoting that at each vertex one $i$-gon, one $j$-gon and one
$k$-gon is joined. The simplest such geometric structure is the $V=8$ cube, with
vertex structure $(4,4,4)$, denoting that three squares are joined at each
vertex. The cube is followed by the $V=24$ truncated octahedron, with vertex-face
structure is  $(4,6,6)$ , denoting that at each vertex one square and two
hexagons are joined. Allowing for octagon, and higher faces lead to only two
more closed Archemedian polyhedra, $V=48$ $(4,6,8)$ and $V=120$ $(4,6,10)$, in which
one square and one hexagon are joined to an 8 or 10 sided polygon at each
vertex. These are the most symmetric  QCD Fullerenes as illustrated in Fig.
~\ref{f:fig2}.

Infinite two dimensional tiling or fences can also be
created, for example 
$(4,8,8)$, $(4,6,12)$ and the J-graphite $(6,6,6)$. 
In addition, as with carbon cages, there are of
course many closed structures with less symmetry such as  junction $n$-prisms
$(4,4,2n)$ that can be constructed.    Here we will not attempt to discuss the
dynamics of elementary particle or heavy ion/nuclear collisions that may lead
to the formation of QCD J-balls. In nuclear collisions, we simply assume that
the observed copious production of  baryons and anti-baryons (interpreted as
junctions and anti-junctions) in central collisions is sufficient to allow
such a configuration to form with finite probability amidst the ``nuclear ashes"
due to its relatively high binding energy.  The carbon Fullerenes 
$C_{60}$ and $C_{70}$
were similarly found within the ashes of laser seared graphite.   Another
mechanism to create QCD Buckyballs may exist also, that has no analogy
in Fullerene chemistry. In particular, high energy collisions of protons and
anti-protons at the Fermilab Tevatron accelerator satisfy the conditions for
zero net baryon number and high energy density, that are required to excite QCD
Fullerenes out of the physical vacuum of strong interactions.

To estimate the relative binding energies of  QCD Fullerenes, 
we consider   the simplest model for the relative energy of 
J-balls consistent with QCD~\cite{jb6}. For a
J-ball consisting of $ V/2$ junctions and $V/2$ 
anti-junctions connected to form a
polyhedron with $E$ edges with lengths $l_i$ we take
the following model Hamiltonian
\begin{equation}
	H(l_i, {\bf n}_{v,i}; V, E) =
		\sum_{i=1}^E	(\frac{a}{l_i} + \kappa l_i) 
		+ \gamma \sum_{v = 1}^V 
			\sum_{i<j = 1}^3 {\bf n}_{v,i} {\bf n}_{v,j}
\end{equation}

where  ${\bf n}_{v,i}$, $i = 1,2,3$  are the three unit vectors
 pointing away tangent to the edges at vertex
$v$. Implicit above is that the topology is defined by these unit vectors and
that the flux tubes are straight lines between vertices. The first term is a
``kinetic" or  ``vertex localization" energy, 
with coefficient $a$ that is not yet
precisely known. However, one can estimate that $a \approx \pi h$ from
$\omega = (2 \pi \hbar)/\lambda$ and assuming that $\lambda = 2 l $
that holds in the case of the lowest excitation for a string with two
fixed ends.
 The second parameter of the effective  
Hamiltonian is the confining string
tension,  $\kappa(T)\approx 1$ GeV/fm, 
a term that vanishes above the deconfinement temperature, $T_c \approx 150$ MeV.
The postulated ``strain'' term with strength $\gamma$
 is analogous to the Biot-Savart
law in circuits and plays the  role of  bond angle strain in carbon
nanostructures.  In this model the relative binding energies are determined by
the last term although its magnitude is not yet known from lattice QCD. 
We estimate below the possible range of $\gamma$ and use these
limiting values to give a semi-classical estimate of the
mass range of the QCD Fullerenes.

 For a vertex and face structure $V$ and $(n_1,  n_2,  n_3)$ 
the total strain energy is
\begin{equation}
	\delta h_V =	\frac{\Delta H_V}{V}  =  - \gamma 
		\left[\cos\left(\frac{2\pi}{n_1}\right) + 
		\cos\left(\frac{2\pi}{n_2}\right) + 
		\cos\left(\frac{2\pi}{n_3}\right)\right]. 
		\label{e:dham}
\end{equation}

 For the $V=8$ J-cube with face structure $(4,4,4)$
and $\Delta h_8 = 0$. For the $V=24, 48, 120$ J-balls , this strain energy
per vertex is $-1$, $-(1 + \sqrt{2})/2 = -1.207$, 
$-(3 + \sqrt{5})/4 = - 1.309$ in units of $\gamma $. 
The absolute minimum is reached for the
junction graphite fence which is bound with $-3/2 \gamma$  
per vertex.  The  $(4,8,8)$ and
$(4,6,12)$ tiles are only bound by $-\sqrt{2}\gamma$  and 
$-0.5(1+\sqrt{3}) \gamma $ per vertex. In contrast, the $ n$ prisms
$(4,4,2 n)$ are bound by $-\cos(\pi/n)> -1$. 
Note that the $(J\underline{J})_1= M_J^0$ 
is most unfavorable due to its maximum strain energy of $6\gamma$. 
The $V=24$  J-ball in Figs.~\ref{f:fig1},\ref{f:fig2} with total strain
energy $-24\gamma$ may be particularly stable not 
only because of its relatively low
strain energy but also due to its topological arrangement of junctions and
anti-junctions that increases the potential barrier between adjacent 
$J\underline{J}$
annihilation. The vertex structure of this $V=24$ QCD Buckyball, 
$(4,6,6)$, is the
closest to that of the carbon Buckyball 
$C_{60}$ whose  vertex structure is $(5,6,6)$.

\subsection{Semiclassical mass estimates for QCD Fullerenes} 
Let us find the semi-classical values of
the Hamiltonian $H$ to give an estimate of the
expected mass range of the $V = 8, 24, 48$ and $120$ QCD
Fullerenes. 
Let us first observe that the minimum of the
$H$ Hamiltonian can be determined from requiring that 
\begin{equation}
	\frac{\partial H}{\partial l_i} =0
\end{equation}
for all $i=1, ... V$, which implies that all the edges have
the same length of 
\begin{equation}
l_i = l_j = l = \sqrt{\frac{a}{\kappa}} \approx  0.79 \,\,\, \mbox{\rm fm}.
\end{equation}
The mass of a QCD Fullerene can be semi-classically approximated
by the value of $H$ at this minimum,
\begin{equation}
	M_V = (\frac{3}{2} \sqrt{a \kappa} + \delta h_V)  V
\end{equation}
Hence the mass of these QCD Fullerenes is always proportional
to the number of vertexes $V$ and the constant of proportionality
is given by a sum of two terms. The first term is a kinetic term,
that can be estimated as $\frac{3}{2} \sqrt{\pi 0.197 }$ GeV 
$\approx 1.18$ GeV, while the second strain term is a product of a
known geometrical contribution and the unknown constant of proportionality
$\gamma$. As $a > m_N= 0.940$ GeV, we find that without a strain term the
mass of QCD Fullerenes were about 25 \% higher than
that of a system consisting from $V/2$ nucleons and $V/2$ anti-nucleons,
hence if $\gamma = 0$ these excitations are most likely unstable.

At what value of $\gamma=\gamma_c$ 
were at least some of the QCD Fullerenes stable? 
Including the possibility of tilings, the absolute minimum of the
geometrical contribution to the strain term is 
$\sum_{i<j=1}^3 {\bf n}_i {\bf n}_j = - 1.5$, that can be achieved within
a graphite like layer. Hence one obtains for the critical value of $\gamma$
\begin{equation}
	\gamma_c = \sqrt{a \kappa} -  \frac{2}{3} m_N
\end{equation}
which leads to a numerical estimate of $\gamma_c \approx 0.16 $ GeV.
If in Nature $\gamma > \gamma_c$ then (at least some high mass)
 QCD Fullerenes are expected to be stable against decay to baryon anti-baryon
pairs, while if $\gamma < \gamma_c$ all these objects are unstable
for such decays and may exist only as short lived resonance states. 

Let us now determine an absolute lower and an upper limit for the strain
coefficent $\gamma$. 
If $\gamma$ were negative, the strings were attracted to
each other and the $M^0_J$ state would be a stable bound state
and it would be difficult to explain why this  state has
not been observed untill now.  Excluding this possibility,
one obtains $0 \le \gamma $. 
An upper limit on the possible values of $\gamma$ can be obtained from
requiring that even for a graphite like tiling the mass of the Fullerene
should be positive, which  implies 
$ \gamma < \sqrt{a \kappa}$ . Thus one obtains the following lower
and upper limits for the mass of QCD Fullerenes:
\begin{equation}
	 \frac{3}{2} V \sqrt{a \kappa}  \le M_V < 
	\left[ \frac{3}{2} V - \cos\left(\frac{2\pi}{n_1}\right) 
		- \cos\left(\frac{2\pi}{n_2}\right) 
		- \cos\left(\frac{2\pi}{n_3}\right)\right] \sqrt{a\kappa}.
\end{equation}

Utilizing these limiting values, we obtain Table \ref{t:1}
that summarizes the mass range estimates for the most symmetric QCD Fullerenes
 utilizing the geometric strain coefficients determined by eq.
~(\ref{e:dham}).
\begin{center}
\begin{table}
\begin{tabular}{ccccccc} \hline\hline 
$V$ 
	& $(n_1,n_2,n_3)$
		& $\delta h/\gamma$
			& $M_{{\rm min}}$ (GeV)
				& $M_{{\rm max}} $ (GeV)
					& $M_{{\rm crit}}$ (GeV) 
						&$ d$ (fm) \cr \hline
8	& (4,4,4)
		& 0	& 9.4 	& 9.4 	& 7.5 	& 1.3   \cr
24	
	& (4,6,6)
		& -1	& 9.4 	& 28.3  & 22.6  & 2.5   \cr
48	
	& (4,6,8)
		& $-\frac{1+\sqrt{2}}{2}$ 
			& 11.1 	& 56.7  & 45.1  & 3.6   \cr
120	
	& (4,6,10)
		& $-\frac{3+\sqrt{5}}{4}$ 
			& 18.0 	& 141.7  & 112.8 & 6.0  \cr \hline\hline 
	&&&&\cr
\end{tabular}
\caption{Estimated mass range for various QCD Fullerenes.
$V$ stands for the number of vertexes, $(n_1,n_2,n_3)$ for
the face structure at a vertex, $M_{\rm min}$  and $M_{\rm max}$
are the estimated lower and upper limits for the mass of the
QCD Fullerene, 
together with
the critical mass of stability $M_{\rm crit}$. 
The diameter of the circumscribed sphere,
$d$ was  estimated from $l \approx 0.79$ fm 
and the geometrical structure. }
\label{t:1}
\end{table}
\end{center}
Although Table~\ref{t:1} contains order of magnitude estimates only,
 we can already  observe interesting patterns. In particular, the
strain coefficient does not influence the mass estimate for the $V=8$
QCD cube, and the estimated mass is much higher than that of 8 nucleons
hence this and all the low lying QCD Fullerenes are expected to be
unstable as they are even more strained than the cube.
The first reasonable candidate would be a $V =B + \overline{B} = 24$
QCD truncated octahedron.  The most stable candidates are expected to be
 the $V = 48$
{\it QCD Great Rombicuboctahedron } and the $V = 120$
{\it QCD Great Rombicosidodecahedron}. These structures 
are compact but less strained than similarly compact lower excitations.
Their compact structure and 
their favourable strain term may stabilize all three of them 
in a large domain of the allowed parameter space.

\section{CP odd J-ball states in QCD}
\label{s:cpodd}
The junction $n$-prisms can be regarded as a closed ribbon of 
$n$ $J\underline{J}$  pairs. Under
simultaneous charge conjugation and parity transformation, these prisms are
invariant and hence CP even as are all the junction 
Fullerenes shown in Fig.~\ref{f:fig2} .
However, other nontrivial topological configurations can be constructed which
are not symmetric under CP. For example an odd number of 
$J\underline{J}$  pairs cannot be
closed into a prism due to the oriented flux at the junctions, but after a
twist to right or left can be connected into a Moebius strip. The two Moebius
strips transform into each other 
and thus there exists a linear combination of
the two that is odd under CP. Hence QCD $J$-ribbons can be characterized by a
single ``winding number" $(i)$ that gives the number of twists before the ribbon
is closed on itself. The topology of the excitations of QCD seems to be  very
interesting, because not only ribbons but also tubes can be formed. The ends
can be closed with caps formed by squares, octagons and decagons, satisfying
eq. (5), or can be open, ending on valence quarks. The QCD femto-tubes are
analogous to the carbon nano-tubes,  both may have interesting chiral
properties. As carbon nano-tubes, the QCD femto-tubes can be characterized by
two integers $(i,j)$, which gives the number of steps in the direction of the
lattice vectors, that connect equivalent points on the surface of the tubes.
Another interesting possibility is to close the 
$J$-tube on itself, creating a
toroidal structure. The femto-tubes can be closed  by connecting the two ends
of a long tube, and these ends can be rotated before the connection. This gives
QCD femto-tori that can be characterized by 3 winding numbers, the $(i,j,k)$
femto-tori.

\section{Summary}
\label{s:summ}
 Fullerene type of pure glue topological configurations can be
constructed in QCD. These ``J-balls" are  QCD femto-structures
with the highest geometrical symmetry. All of the QCD Fullerenes  have an equal
number of junctions and anti-junctions, and may have specific  geometrical and
topological properties.  The QCD Buckyballs are CP even, other QCD structures
such as linear combinations J-Moebius ribbons can be constructed that are CP
odd. Topological winding  numbers can be introduced to characterize these
states. The QCD femto-ribbons are characterized by a single integer  $(i)$, the
femto-tubes by a pair of integers $(i,j)$, while the QCD femto-tori by a triplet
of integers, $(i,j,k)$.   We determined that the most symmetric (likely most
stable)  QCD Buckyball configurations have the magic numbers of baryons +
anti-baryons $B + \overline{B} = 8$, $24$, $48$ and $120$. 
Although these configurations are
likely unstable, they are expected to be more stable than clusters of baryons
and anti-baryons with different junction numbers, and they may  appear as peaks
in the spectrum of  $(B\overline{B})_n$ clusters 
with a given total baryon+antibaryon number.
To create them, high initial energy densities and small net initial baryon
number densities and large volumes are needed.  Such conditions may exist in
the mid-rapidity domain of central $Au+Au$ collisions   at RHIC or LHC as
well as in diffractive collisions of protons and anti-protons at the Fermilab
Tevatron accelerators.  We suggest  to search for clusters of  baryons and
anti-baryons with multiparticle correlation patterns of the vertices of J-balls
in rapidity slices. In addition, searches for CP violating domains at RHIC
should look for unusual baryon anti-baryon correlations suggested by our
J-Moebii structures. Baryon junction and anti-junction networks may also help
to understand the structure of domain walls between different CP vacua in QCD.

\section*{Acknowledgments}
One of us (T. Cs.) would like to thank M. Albrow and G. Gustafson for
stimulating discussions. This research has been supported by  Hungarian OTKA
grants T026435, T034296, the Dutch-Hungarian NWO-OTKA grant N025487, 
the Collegium Budapest,  an US - Hungarian NSF - MTA - OTKA grant, 
by the Brazilian FAPESP and by the US Department of
Energy DE-FG02-93ER40764.

\vfill\eject
\end{document}